\documentclass[5p,authoryear]{elsarticle}
\usepackage[utf8]{inputenc}
\usepackage{amssymb}
\usepackage{amsmath}
\usepackage{url}
\usepackage{caption}
\usepackage{subcaption}
\usepackage{graphicx}
\usepackage{color}
\usepackage{microtype}

\usepackage{booktabs}

\usepackage{listings}
\usepackage{color}

\journal{astro-ph}

\begin{document}

\newcommand{\code}[1]{\texttt{#1}}
\newcommand{\degrees}[1]{$#1^\circ$}

\newcommand{\githash}{f893517}\newcommand{\gitdate}{2016-06-12}

\begin{frontmatter}

\title{The 4 Pi Sky Transient Alerts Hub \tnoteref{git}}
\tnotetext[git]{Generated from git source \texttt{\githash} dated \gitdate.}

\author{Tim~D.~Staley}
\ead{tim.staley@physics.ox.ac.uk}
\author{Rob~Fender}
\address{Astrophysics, Department of Physics, University of Oxford, 
Keble Road, Oxford OX1 3RH, UK}

\begin{abstract}
We introduce the 4 Pi Sky `hub', a collection of open data-services and underlying software packages built for rapid, fully automated reporting and response to astronomical transient alerts. 
These packages build on the mature `VOEvent' standardized message-format, and aim to provide a decentralized and open infrastructure for handling transient alerts. 
In particular we draw attention to the initial release of `\texttt{voeventdb}', an archive and remote-query service that allows astronomers to make historical queries about transient alerts. 
By employing spatial filters and web-of-citation lookups, \texttt{voeventdb} enables cross-matching of transient alerts to bring together data from multiple sources, as well as providing a point of reference when planning new follow-up campaigns.
We also highlight the recent addition of optical-transient feeds from the ASASSN and GAIA projects to our VOEvent distribution stream.
Both the source-code and deployment-scripts which implement these services are freely available and permissively licensed, with the intention that other teams may use them to implement local or project-specific VOEvent archives. 
In the course of describing these packages we provide a basic primer for getting started with automated transient astronomy, including a condensed introduction to the VOEvent standard.
\end{abstract}


\end{frontmatter}

\section{Introduction}
\label{sec:intro}
For over two decades, real-time transient alert networks such as NASA-GCN have enabled ground-breaking astronomical results. 
The late 90's gave us rapid optical localisation and redshift determination of gamma-ray bursts (GRBs), pinning down their extra-galactic origins and extraordinary levels of luminosity. The intervening years have seen a wealth of science from short-and-long GRBs, flare stars, magnetars and more \citep[e.g.][]{Gehrels2015}. 
Rapid follow-up on timescales of hours-to-days has been crucial for characterising many of these events. 
More recently, exciting results such as the first gravitational wave detection \citep{Abbott2016}, and possible localisation of a fast radio burst \citep{Keane2016}, have added to the classes of astronomical events which benefit from rapid alerts and response.

However, follow-up and discovery announcements from the wider astronomical community are still 
a largely manual affair. 
Almost all ground-based transient detections are reported via plain-text messages, posted to
Astronomer's Telegram, GCN-circular notices, the Central Bureau for Astronomical Telegrams, and similar. 
This model has undoubtedly served the community well for many years, and we expect this practice to continue indefinitely for transients of note. 
Nonetheless, wider automation of transient reporting and response would allow for more effective early identification of rare and interesting (or simply rapidly evolving) events.
Increasing the frequency and variety of machine-readable transient reports would enable much greater sophistication and co-ordination in automated follow-up campaigns.
Additionally, we expect that within the next decade a more automated approach will become essential to prevent follow-up projects from being overwhelmed as the rate of reported transients increases
with surveys such as those undertaken by the SKA and LSST \citep{Fender2015a,LSST2009}.

None of these arguments are new ---
astronomers have been exploring approaches to large-scale automation of transient-astronomy and alert collation for over a decade \citep[e.g.][]{Allan2004, Williams2010}.
However, we contend that influential external factors have changed in the intervening years.
We are now definitively seeing the rise of `big data' astronomy; huge survey projects with accordingly large datasets and data-analysis teams. 
GAIA \citep{Perryman2001} is a current example that is now actively providing transient alerts, while SKA and the LSST loom in the near-future, promising to bring vastly increased rates of transient discovery. 
Simultaneously, from the `bottom-up' perspective we are seeing an increase in levels of code-sharing and reuse in astronomy, as exemplified by projects such as \texttt{Astropy} \citep{Astropy2013} and the increasingly large number of codes listed in the Astrophysics Source Code Library  \citep{Allen2015}. 
This change in culture greatly improves the chance of long-term success for open-source, community maintained software packages that make getting started with automated transient astronomy a more straightforward and scientifically productive affair.
We hope that the combination of these effects will finally tip the effort/reward trade-off in favour of a more automated approach to transient follow-up for the average astronomer.

Since 2012 we have been developing various tools for working with astronomical transient alerts distributed via the VOEvent standard, which provides a mature protocol and format for automated reporting of astronomical transients. 
Development was initially for our own research \citep[and has contributed to a number of scientific results, e.g.][]{Anderson2014,Fender2015b}, but the software packages have now reached a level of functionality and maturity that we judge fit for public release.

The rest of this paper provides an introduction to the services available via 4 Pi Sky, and the accompanying software packages developed. 
We aim to provide a stable, easily located service supplying a consolidated stream of all transient alerts which are publicly available in machine-readable format (colloquially, a `hub'). 
We note that while a single point of access for a data-stream is convenient, it also represents a single point of failure.
Our hope is that we may build some level of confidence in the system by open-sourcing and documenting the underlying infrastructure that runs the 4 Pi Sky hub, since other teams can then replicate the functionality and provide services of their own with relative ease (and we encourage them to do so).

\begin{figure*}[htp]
\begin{center}  
  \includegraphics[width=.96\textwidth]{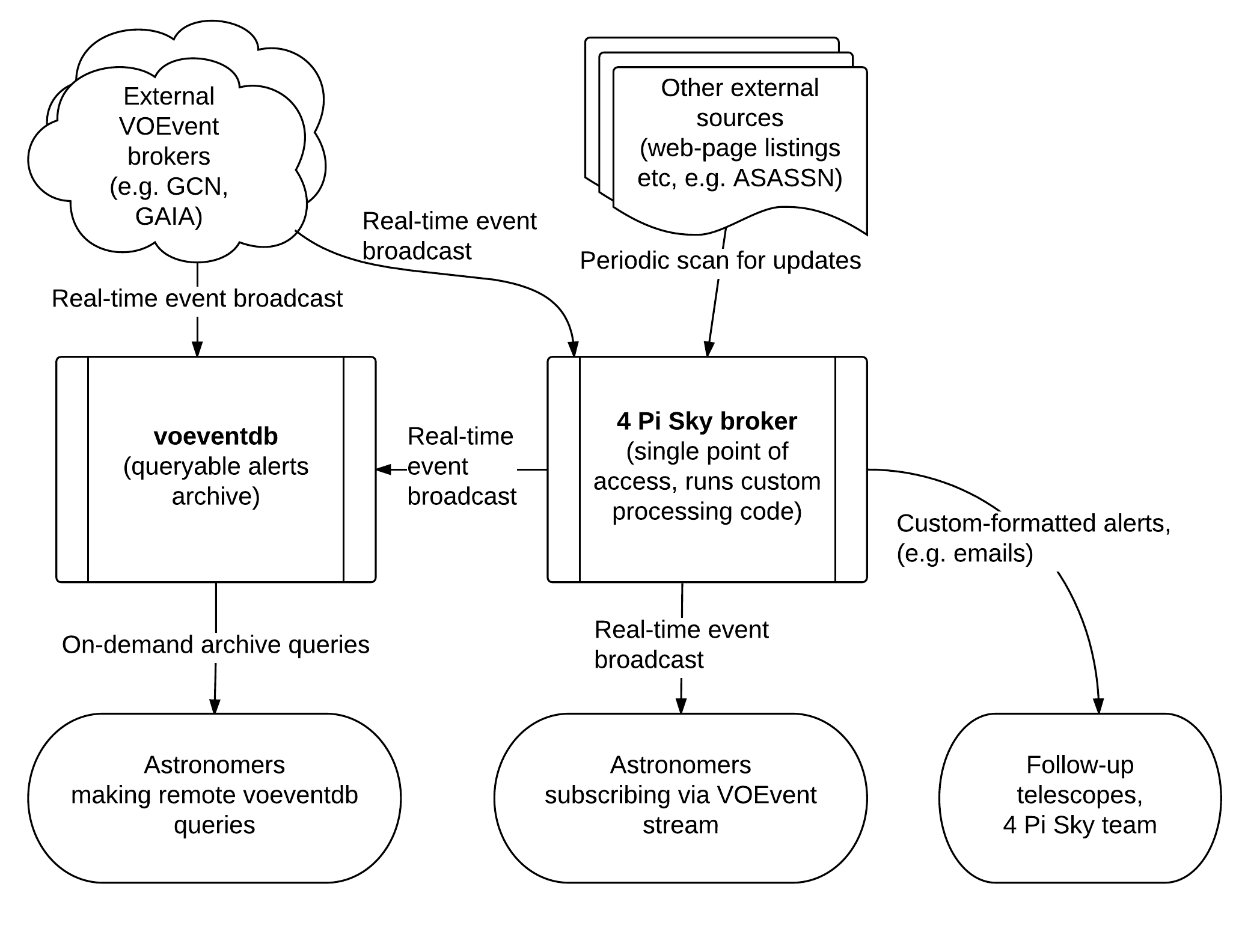}
  \caption[Dataflow overview for the 4 Pi Sky hub]{%
  Data-flow overview for the 4 Pi Sky hub. See text for details. 
  Figure~\ref{fig:dataflow-detail} depicts the same system broken down into its constituent software components.
\label{fig:dataflow-overview}
} 
\end{center} 
\end{figure*}

We introduce the services and software by outlining features and functionality, deferring detailed technical description and instructions to their respective documentation.
End-user tools are described first, proceeding to the underlying implementations which may be of interest to those wishing to provide services of their own. 
Figure~\ref{fig:dataflow-overview} gives a high-level overview of the dataflow through the 4 Pi Sky hub, and may be a useful point of reference for the following sections.

\section{The VOEvent standard (a \textit{very short} introduction)}
The 4 Pi Sky software packages make extensive use of the VOEvent (Virtual Observatory Event) standard \citep{Seaman2008,Seaman2011}, which encodes data using XML (Extensible Markup Language). 
We give a brief introduction to the standard for unfamiliar readers, hopefully giving just enough context to follow the rest of this paper.
For a comprehensive introduction to VOEvents we refer the reader to \citet{Swinbank2015}\footnote{%
\url{http://voevent.readthedocs.org/}
}.

The VOEvent standard can be briefly introduced by outlining the contents of the permitted sections:
\begin{description}
  \item[\textit{(Header)}] \hfill \\
  Every VOEvent is prefaced by a header section. 
  This contains a unique identifying string, known as an \textit{IVORN}, and also states the role of the VOEvent, which can be one of three options: `observation', `utility', or `test'.
  \item[\textit{Who}] \hfill \\
  The author(s) of the VOEvent, and when the VOEvent was created.
  \item[\textit{What}] \hfill \\
  Various details of the event (e.g. parameters specific to the particular observatory and/or event-type).
  \item[\textit{WhereWhen}] \hfill \\
  Localisation of the event, and when it was observed.
  \item[\textit{How}] \hfill \\
  How the event was observed (typically contains a reference to a detailed instrument description).
  \item[\textit{Why}] \hfill \\
  Suspected physical cause (or multiple possible causes) of the event, e.g. whether brightening is due to a supernova or a gamma-ray burst.
  \item[\textit{Citations}]
  References to earlier VOEvents. 
  This section allows for linking follow-up observations to an original discovery; where necessary it can also be employed to issue retractions.
\end{description}
Most of the sections follow a restricted format, and can usually be interpreted in a uniform manner.
The exception is the \textit{What} section, which takes a flexible `key-value pair' format, with author-chosen keys and optional grouping of entries. 
This is generally used for project-specific parameters which do not fit neatly into any of the predefined sections.

\subsection{The \texttt{voevent-parse} package}
Due to the choice of XML as the underlying data-structure, the VOEvent standard benefits from a selection of widely available, mature software libraries for manipulating and validating data-packets according to the VOEvent data-schema.
However, general-purpose XML libraries are necessarily flexible and many-featured, and using them effectively can be complex. 
For this reason we developed \texttt{voevent-parse}, 
a Python package which provides a convenient interface to the \texttt{lxml} library, aiming to simplify many common tasks related to manipulating VOEvents \citep{Staley2014}. 
The \texttt{voevent-parse} package is accompanied by a full set of documentation and introductory tutorials.\footnote{\url{https://github.com/timstaley/voevent-parse}}

\section{Receiving and distributing VOEvents via the 4 Pi Sky VOEvent Broker}
\subsection{What is a VOEvent Broker?}
While the VOEvent standard defines a data format, it makes no assertions as to how VOEvents should be transmitted --- VOEvent packets can be emailed, uploaded to a web-page, or printed and sent by carrier pigeon \citep{Waitzman1990}. 
Obviously, for transient follow-up applications a low-latency method is preferred. 
One such method is the VOEvent Transport Protocol \citep[][]{Allan2016}, as implemented by software packages such as \texttt{Dakota}\footnote{\url{http://voevent.dc3.com/}} \citep{Denny2010} and \texttt{Comet}\footnote{\url{http://comet.readthedocs.org/}} \citep{Swinbank2014}. 
The VOEvent Transport Protocol enables creation of a distributed network of participants, defining three key roles \citep[cf][]{Swinbank2014}:
\begin{description}
  \item[\textit{Author}] \hfill \\
    An author is responsible for creating and publishing one or more VOEvents.
    \item[\textit{Subscriber}] \hfill \\
    A subscriber receives the VOEvents generated by one or more authors.
    \item[\textit{Broker}] \hfill \\
    A broker receives VOEvents from other network entities and re-distributes them to one or more subscribers. In addition, a broker may perform “added value” services. These could be at the request of particular subscribers (e.g. to apply a filter to the event stream sent to that subscriber), or applied more generally to the event stream (e.g. to apply some annotation to all events processed).
\end{description}

\subsection{VOEvent streams currently available via 4 Pi Sky}
We maintain an open-access VOEvent broker, powered by \texttt{Comet}, accessible via \\\mbox{\url{voevent.4pisky.org}}.
This represents an accessible initial point-of-access for real-time alerts --- those interested can install the \texttt{Comet} Python package, connect and start receiving VOEvents from multiple observatories immediately. 
Currently we redistribute alerts released in real-time or near-real-time from:
\begin{itemize}
 \item \textbf{NASA-GCN}, e.g. alerts from the \textit{Swift}, \textit{Fermi} and \textit{INTEGRAL} satellites.
 \item \textbf{GAIA photometric alerts}, providing a real-time feed of the alerts concurrently published via the GAIA science alerts website\footnote{\url{http://gsaweb.ast.cam.ac.uk/alerts}}.
 \item \textbf{ASASSN} \citep{Shappee2014}, as published via the ASASSN alerts web-page\footnote{\url{http://www.astronomy.ohio-state.edu/~assassin/transients.html}}.
 \item 4 Pi Sky \textbf{ALARRM} alerts, which announce when a request for radio follow-up has been triggered as part of our own transients follow-up programme.
\end{itemize}

\section{Processing and sending VOEvents}
\subsection{fourpiskytools}
The next logical question is then `how do I do something useful with the VOEvents I receive?'
Rather than attempt to build a standard tool that satisfies the myriad filtering-and-response requirements of the astronomy community, we instead encourage end-users to implement processing scripts of their own. 
To this end we have published a small collection of code examples under the moniker `\texttt{fourpiskytools}'\footnote{\url{https://github.com/4pisky/fourpiskytools}}. 
This repository contains an installation walkthrough and some minimum working examples,  demonstrating how to configure \texttt{Comet} to receive VOEvents from the 4 Pi Sky broker, pipe them to a processing script, and (in the provided example) send a notification to the desktop every time a VOEvent is received, with optional filtering.
The repository also contains some basic routines for generating unique ID stamps and sending newly authored VOEvents to a broker, which should prove useful to those interested in publishing their own events.

\subsection{fourpisky-core}
The codebase we use for processing and triggering VOEvents for our own scientific programs, `\texttt{fourpisky-core}', is significantly more complex than the examples provided under the \texttt{fourpiskytools} package, and largely undocumented. 
Nevertheless, it may still prove a useful reference for experienced Python programmers, and is available for inspection.\footnote{\url{https://github.com/4pisky/fourpisky-core}}
Additionally, making the codebase open makes it possible for third parties to inspect the logic behind the the sending of any new triggers from our broker.

Functionally, the codebase plays two roles. 
The first is to identify VOEvents of interest, calculate visibility schedules for various observatories, and send out observation-requests and notifications as appropriate. 
Timely event processing is achieved through use of the \texttt{Celery}\footnote{\url{http://www.celeryproject.org/}} task queue framework; with this approach we are able to process a sustained rate of fifteen events per second on a single-core virtual machine with only 512MB of RAM. 
Bursts of VOEvents received at higher rates are simply queued and processed as soon as possible.
This implies that we can process of order one million events per day with infrastructure costs that are practically negligible, assuming VOEvents of size and complexity similar to those seen to date.
In addition, it is trivial to scale up to higher processing rates, as the \texttt{Celery} framework allows task-processing distribution over multiple CPU cores and even multiple machines.

The second duty performed by \texttt{fourpisky-core} is that of a feed-in, or `web-scraping' role. 
This provides a means of accessing machine-readable transient alerts (typically in the form of an HTML table or CSV file) which are publicly available via the web, converting them to VOEvent and then broadcasting them via the VOEvent network.
We have developed a few core routines to handle fetching data, checking for updates using minimal bandwidth, etc. 
These routines are run automatically on a regular basis, which allows for redistribution and archiving of transient alerts from sources which might not otherwise be available in uniform VOEvent format (for example, the ASASSN alerts feed).

\section{Accessing and querying historical VOEvents}
The VOEvent standard provides a widely applicable standard for distributing transient alerts, but to date there have been no services for programmatically searching through historical events.
We have implemented an archival service, `\texttt{voeventdb}', with an open API that allows for retrieval and complex querying of historical VOEvents. 
This serves key roles in a transient alerts network. 
Firstly, it allows people distributing or monitoring VOEvent alerts to recover any missed real-time events, in the event of a network or systems outage. 
More generally, it allows astronomers to search through the archive of VOEvents, which can be useful for estimating event rates when planning future observations, looking for related events in a particular region of sky, or mapping the distribution of detected events, etc.

The \texttt{voeventdb} codebase is split into two packages, \texttt{voeventdb.server} and \texttt{voeventdb.remote}.

\subsection{\texttt{voeventdb.server}}
The \texttt{voeventdb.server}\footnote{\url{https://github.com/timstaley/voeventdb}}
package implements both the database store for archiving VOEvent packets, and the HTTP interface for performing remote queries on the database. 
The database schema is quite minimal, aiming to represent the basic information we expect to be present in most if not all VOEvent packets - IVORN, author information and timestamp, role, etc. 
Separate tables contain linked entries for sky-event details and inter-VOEvent citations, where present. 
A key feature is that the full XML content of every VOEvent is stored and easily retrievable. 
This means that any project-specific parameters in a VOEvent can easily be accessed by retrieving the relevant XML data and processing locally. 
Similarly, any VOEvent processing code can be applied or tested on either VOEvents received in real-time via the broker network, or archival VOEvent packets retrieved from the archive, since the format is identical.
The server codebase is built using the \texttt{Flask}\footnote{\url{http://flask.pocoo.org/}} web framework and makes extensive use of the \texttt{SQLAlchemy}\footnote{\url{http://www.sqlalchemy.org/}} SQL toolkit for handling database queries. 
The different HTTP data access points (AKA `views' or `endpoints') and the various applicable query-filters are implemented using a handful of base classes, with each child-class specifying only the view-name, relevant SQL query, and any processing steps required to display the data. 
This results in a relatively compact codebase which is easy to extend by adding additional viewpoints or filtering capabilities.
The combinations of data-viewpoints and query-filters are extensively tested with an automated test-suite powered by the \texttt{pytest}\footnote{\url{http://pytest.org/}} framework.

\subsection{\texttt{voeventdb.remote}}

\lstset{ %
language=Python,                
basicstyle=\sffamily,
columns=fullflexible,
backgroundcolor=\color{white},  
showspaces=false,               
showstringspaces=false,         
showtabs=false,                 
frame=single,           
tabsize=2,          
captionpos=b,           
breaklines=true,        
breakatwhitespace=false,    
escapeinside={\%*}{*)},          
caption={%
  Example usage of \texttt{voeventdb.remote} --- see text for details.},
}
\begin{lstlisting}[float,label=snippet]
import voeventdb.remote.apiv1 as apiv1
from astropy.coordinates import Angle, SkyCoord

# Define a search cone
cone_centre = SkyCoord(ra=149, dec=51, unit='deg')
cone_radius = Angle(3, unit='deg')

# Define a filter-set
my_filters = {
    apiv1.FilterKeys.cone: (cone_centre, cone_radius),
    apiv1.FilterKeys.ivorn_contains: 'BAT_GRB',
    apiv1.FilterKeys.role: 'observation',
}

# See how many results match our filter-set
n_results = apiv1.count(filters=my_filters)

# List all IVORNs for matching VOEvents, oldest first
matching_ivorns = apiv1.list_ivorn(
    filters=my_filters,
    order=apiv1.OrderValues.author_datetime,
)
# Fetch the raw XML for the first listed VOEvent
xml_content = apiv1.packet_xml(matching_ivorns[0])

\end{lstlisting}

For most users of the VOEvent archive service, we expect ease-of-use to be a primary concern. 
With this in mind, we have implemented a Python client library, \texttt{voeventdb.remote},\footnote{\url{https://github.com/timstaley/voeventdb.remote}}
which provides a relatively straightforward means of querying the archive. 
Installation should be trivial for most users via the usual Python package installation methods.
The client-library represents each data-view as a Python function, with query-filters represented by Python dictionary structures.
Details such as retrieving large data-lists in paginated chunks are taken care of, so the end-user simply sees the concatenated results. The package is accompanied by a full set of documentation and tutorial notebooks.

Listing~\ref{snippet} illustrates use of the \texttt{voeventdb.remote} library.
The listing first defines a search-cone of radius three degrees about the given co-ordinates. 
We then use that search-cone along with the requirements that the IVORN should contain the substring `BAT\_GRB', and that the VOEvent role should be `observation', to define a set of filters limiting the search to \textit{Swift}-BAT GRB alerts within this cone. 
We then check the number of matching results and fetch a list of matching IVORNs.
Finally, the first entry in the list of matching IVORNs is used to fetch the full XML-content of the corresponding archived VOEvent.

\subsection{Interactive usage and querying from the browser}
The \texttt{voeventdb.server} HTTP interface can be viewed and queried manually via a web-browser, but the basic interface is intentionally minimal and really only intended to aid testing and development of client-libraries.
In future we (or others) may consider implementing a user-friendly, interactive graphical web-interface built using the basic HTTP interface as a data-source. 
However, we note that the \texttt{Jupyter}\footnote{\url{https://jupyter.org/}} Notebook platform provides an interactive Python environment which can be used with the \texttt{voeventdb.remote} library to perform queries in an interactive, nonlinear fashion. 
This perhaps reduces the need for a dedicated web-client, which would otherwise be a burden in terms of added development and maintenance effort, compared to maintaining a sole Python-library. 

\begin{sloppypar}
 For those who wish to experiment or perform occasional queries on \texttt{voeventdb} without requiring any local installation, we draw attention to the \texttt{voeventdb-dashboard-notebooks}\footnote{\url{https://github.com/4pisky/voeventdb-dashboard-notebooks}} package. 
This is a collection of \texttt{Jupyter} notebooks which have been pre-prepared to run on the \texttt{mybinder}\footnote{\url{http://mybinder.org/}} platform. 
By following links from the Github repository page a user can edit and run ready-made queries using the \texttt{voeventdb.remote} library \emph{from their browser} without any alterations to their local system, albeit with slightly slower performance compared to running a \texttt{Jupyter} notebook environment locally.
\end{sloppypar}

\begin{figure*}[htp]
\begin{center}  
  \includegraphics[width=.96\textwidth]{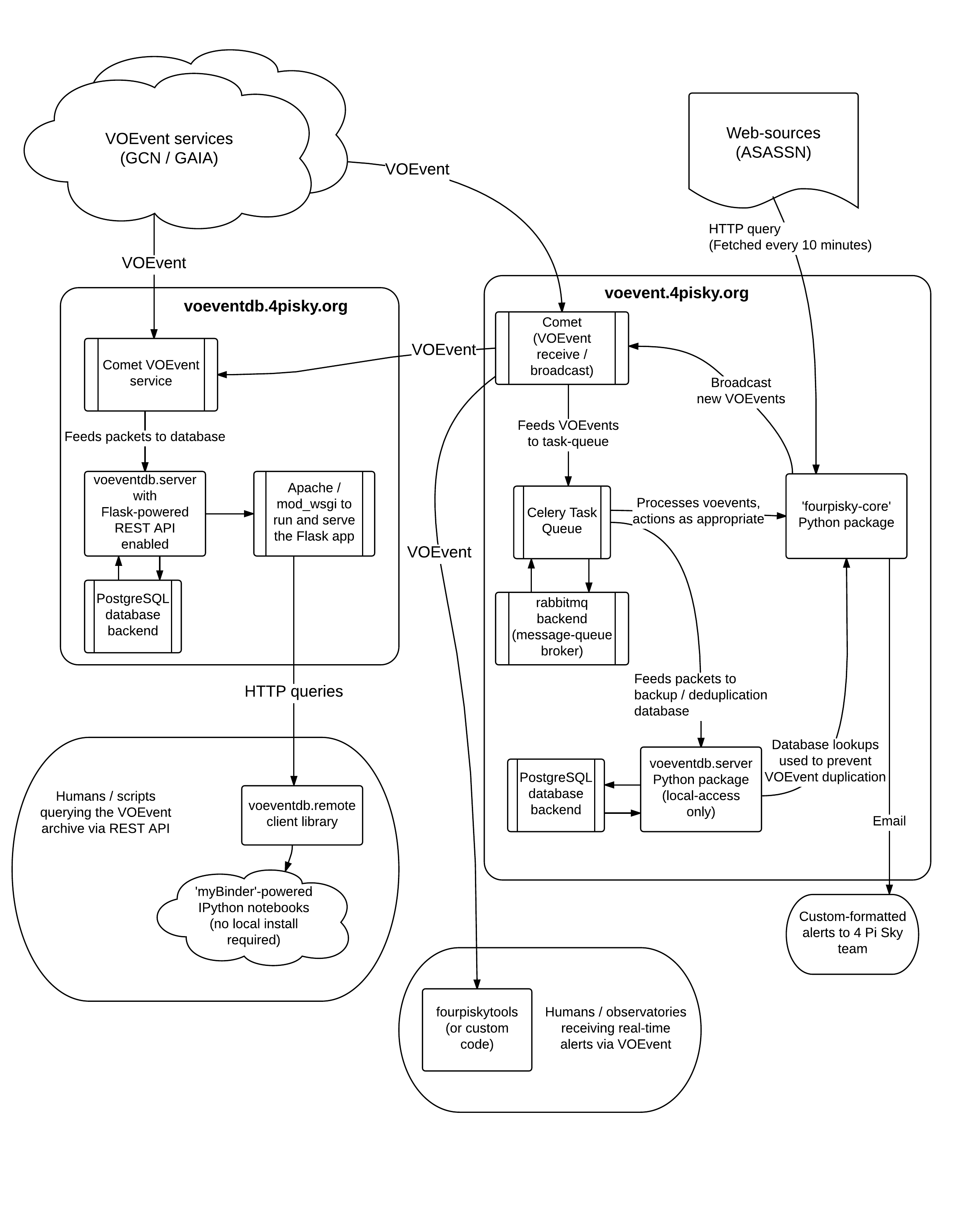}
  \caption[Dataflow and components of the 4 Pi Sky hub]{%
  This figure depicts the same system as Figure~\ref{fig:dataflow-overview}, but with an emphasis on enumerating the major software components that make up the system and the data-formats transmitted between them. 
  Each component requires careful configuration for the system to work as a whole, leading to complex set-up procedures that are best fully automated to achieve reasonable levels of reliability and ease-of-maintenance.
  We describe how we have automated the installation and configuration process in Section~\ref{sec:deployment}.
\label{fig:dataflow-detail}
} 
\end{center} 
\end{figure*}

\section{Open-infrastructure and reproducible deployments}
\label{sec:deployment}

A well-tested and version controlled codebase is necessary, but not sufficient, to run a data-intensive web-service effectively. 
Successful deployment also requires careful configuration of requisite databases, HTTP-servers, and any additional components such as task-queue processors. 
Figure~\ref{fig:dataflow-detail} illustrates this multi-component interaction, expanding on Figure~\ref{fig:dataflow-overview} with emphasis on the software-components and data-formats employed.
Connecting multiple disparate components in this manner is the only sensible way to build a modern data-service --- it's a case of using the right tool for each job, taking advantage of many years of specialized development in each area. 
However, it introduces a host of potential problems due to the added complexity. 
First, the system can only be considered thoroughly tested once it has been installed and configured \textit{as it will be in production}, since some issues may arise due to e.g. differing handling of HTTP queries between a development-and-debugging-server (as provided by most web-development frameworks) and a production HTTP-server like \texttt{Apache}. 
Similarly, care must be taken to produce a consistent installation each time a new system is brought on-line, including easily forgotten details such as database configuration settings, user-permissions, etc.
It also means that any software upgrades must be undertaken with great care to ensure that all components are updated and restarted as appropriate.
These issues are compounded in academia by the fact that most research groups do not have a dedicated system administrator, and so these challenges may be entirely new to the person undertaking the installation.

The difficulties described above point to an obvious solution: automated installation scripts and testing in a `staging' environment (i.e. a full replication of the final configuration) prior to final deployment to production. 
This practice has become widespread in commercial software development in recent years, in tandem with tooling which makes the process simpler and more reliable. 
We have taken advantage of these tools to create a fully scripted staging and deployment process for the 4 Pi Sky data-services, which (crucially) should be flexible enough for reuse by other teams. 

Our deployment scripts make use of \texttt{Ansible}\footnote{\url{https://www.ansible.com}}, which provides a convenient scripting language for system configuration building on a number of core `modules' for common tasks such as installing system packages, downloading or updating a source-code repository, creating a database, etc.
\texttt{Ansible} scripts can be executed against a remote machine (or even multiple remotes in parallel) via SSH.
By design the scripts and modules are intended to be `idempotent', meaning the same deployment scripts can be safely re-run multiple times and used either for initial system installation or for applying later updates. 
The deployment scripts are accompanied by a \texttt{Vagrant}\footnote{\url{https://www.vagrantup.com/}} configuration file for a virtual machine. 
This provides a means of testing the deployment scripts without requiring additional hardware, and can also be utilised as the staging environment for final testing of the system.

Writing effective deployment scripts can be surprisingly complex, although the initial effort is usually repaid many times over by reduced time spent on maintenance and debugging throughout a system's lifetime. 
For complex multi-component systems it makes sense to develop deployment scripts that are separated into corresponding modules, potentially enabling their re-use wherever a given component is required. 
\texttt{Ansible} encourages this practice through a convention of structuring deployment scripts into `roles', scripting modules that provide a given service or perform a complex configuration task. 
Open-source development and sharing of roles is encouraged. 
We have developed roles for installation and configuration of the \texttt{Comet} VOEvent broker\footnote{\url{https://github.com/timstaley/ansible-comet}}, and for the \texttt{voeventdb} archive\footnote{\url{https://github.com/timstaley/ansible-voeventdb}}. 
These are combined to deploy the service at \url{voeventdb.4pisky.org}, and the final deployment scripts are openly available via the 4 Pi Sky Github repository\footnote{\url{https://github.com/4pisky/4pisky-voeventdb}}. 
We hope that these releases prove useful to other teams. 
More generally, we encourage wider consideration of deployment challenges as an integral part of developing and publishing reproducible scientific software packages.

\section{Summary}
We have introduced a collection of software packages and data-services which contribute key functionality to the open-source VOEvent software ecosystem. The packages \texttt{voevent-parse} and \texttt{fourpiskytools}, together with existing VOEvent broker tools, make it simpler to begin working with real-time VOEvent streams. The 4 Pi Sky VOEvents broker provides a collated source of VOEvent streams, distributing alerts from multiple observatories. The newly released \texttt{voeventdb} archive and associated query tools make it possible to access historical alerts, providing crucial cross-matching, testing, and redundancy features.

The software packages are accompanied by documentation, working examples, and (where appropriate) full testing-and-deployment scripts. We hope that they will see uptake and re-use among the community as part of a wider move to automated and `data-driven' time-domain astronomy. At a minimum, the systems described here serve as a proof-of-concept, demonstrating a working end-to-end infrastructure for automated creation, distribution, handling and archiving of astronomical transient alerts.

Finally, we note that the software packages described herein are likely to develop and evolve as time progresses, and so we encourage future readers to consult the VOEvent pages of the 4 Pi Sky website for up-to-date information.\footnote{\url{https://4pisky.org/voevents/}}

\section*{Acknowledgements}

We would like to thank the ASASSN, GAIA science alerts, and NASA-GCN teams for making their transient alerts freely available on short timescales.

We acknowledge ESA Gaia, DPAC and the Photometric Science Alerts Team (http://gaia.ac.uk/selected-gaia-science-alerts).

This work made use of a number of publicly available software packages: Astropy, a community-developed core Python package for Astronomy \citep{Astropy2013}; the IPython package \citep{Perez2007}; matplotlib, a Python library for publication quality graphics \citep{Hunter2007}; and SciPy \citep{jones_scipy_2001}. 

The 4 Pi Sky project was wholly funded by European Research Council Advanced Grant 267697 
`4 $\pi$ sky: Extreme Astrophysics with Revolutionary Radio Telescopes.’

\section*{References}

\bibliographystyle{elsarticle-harv-max8}
\bibliography{fourpisky_hub_intro}

\end{document}